\documentclass[superscriptaddress, nofootinbib, preprintnumbers, twocolumn]{revtex4}

\usepackage{amsmath}	
\usepackage{amsthm}		
\usepackage{amssymb}	
\usepackage{eufrak}
\usepackage{datetime}
\usepackage{graphicx}
\usepackage{xcolor}
\usepackage{verbatim}
\usepackage{bm}
\usepackage{float}
\usepackage{braket}
\usepackage[colorlinks=true, citecolor=midblue, linkcolor=midblue, urlcolor=midblue]{hyperref}
\usepackage{orcidlink}
\usepackage{setspace}

\usepackage{amsmath}	
\usepackage{amsthm}		
\usepackage{amssymb}	
\usepackage{eufrak}
\usepackage{datetime}
\usepackage{graphicx}
\usepackage{xcolor}
\usepackage{verbatim}
\usepackage{bm}
\usepackage{float}
\usepackage{todonotes}

\usepackage[colorlinks=true, citecolor=midblue, linkcolor=midblue, urlcolor=midblue]{hyperref}

\usepackage{setspace}

\definecolor{grey}{rgb}{0.4,0.4,0.4}
\definecolor{dullmagenta}{rgb}{0.4,0,0.4}
\definecolor{darkblue}{rgb}{0,0,0.4}
\definecolor{midblue}{rgb}{0,0,0.5}
\definecolor{midred}{rgb}{0.5,0,0}
\definecolor{orange}{rgb}{1,0.5,0}
\definecolor{lightbrown}{rgb}{0.75,0.5,0.25}
\definecolor{tan}{cmyk}{0.14,0.42,0.56,0}
\definecolor{djunglegreen}{cmyk}{0.99,0,0.52,0}
\definecolor{lightgreen}{rgb}{0,1,0}
\definecolor{olivegreen}{cmyk}{0.64,0,0.95,0.40}
\definecolor{midgreen}{rgb}{0.0,0.675,0.0}
\definecolor{darkgreen}{rgb}{0,0.5,0}

\smallskipamount

\settimeformat{ampmtime}

\usepackage{float}

\begin{document}

\title{Neutrino Masses and Phenomenology in Nnaturalness}
\author{Manuel Ettengruber\,\orcidlink{0000-0001-7331-6370}}
\email{manuel@mpp.mpg.de}
\affiliation{Universite Paris-Saclay, CNRS, CEA, Institut de Physique Theorique, 91191, Gif-sur-Yvette, France
}

\date{\formatdate{\day}{\month}{\year}}

\begin{abstract}
In this paper, it is shown that $N$naturalness scenarios share the intrinsic mechanism to suppress neutrino masses with other infrared neutrino mass models. In such models like extra-dimensional theories or many species theories, the large number of mixing partners is responsible for the neutrino mass suppression. It is shown how neutrino mass matrices arise in $N$naturalness models and the resulting neutrino mixing is analyzed. The first result is that a totally democratic coupling among the different sectors like in the original models is already ruled out by the fact that neutrinos are not massless. In the case where the sector couplings deviate from the intersector ones a tower of additional neutrino mass eigenstates appears whose difference between the squared masses, $\Delta m_{ij}^2$,  is determined fully by the theory. The resulting phenomenology of such a tower is investigated and the unique signals in neutrino oscillation experiments, neutrino mass measurements, and neutrinoless double beta decay experiments are discussed. This opens the door for terrestrial tests of $N$naturalness whose phenomenology was so far focused on Cosmology.

\end{abstract}

\maketitle

\section{Introduction}
Even after decades of intensive research, neutrinos are still particles that did not reveal all their mysteries. On the one hand, they are so well understood that they play an important part in modeling physical processes like supernovae. On the other hand, the basic properties of the neutrinos remain unknown, such as the nature of their mass or even the mass value itself. 

The second particle in the Standard Model (SM) that remains mysterious is the Higgs. Even though the mass of the Higgs is measured, it remains an open question how its mass could be of the order of the weak scale even though a fundamental scalar should feel the physics lying in the UV. This is commonly referred to as the Hierarchy Problem. 

Theories that solve both of these problems simultaneously become particularly intriguing and one of the earliest examples of this is a supersymmetric SO(10) grand unified theory \cite{Clark:1982ai, Aulakh:1982sw, Dvali:1996wh, Brahmachari:1997cq, Aulakh:2000sn, Bajc:2001fe, Aulakh:2003kg}. These theories use supersymmetry to solve the hierarchy problem and generate a neutrino mass term via a Weinberg operator suppressed by a large cutoff scale \cite{Weinberg:1979sa} realized via the celebrated Seesaw mechanism \cite{Minkowski:1977sc, Gell-Mann:1979vob, Yanagida:1980xy, Mohapatra:1979ia, Mohapatra:2004zh}. Up to now, the Seesaw mechanism remains the most popular way to give the neutrino its mass. The key property the Seesaw mechanism relies on is that particles living in the UV suppress the mass of the neutrino. Therefore, one is tempted to call such explanations of neutrino masses "UV solutions". 

Historically the development of the Seesaw aligns with the expectation of physicists that the particles of new physics should live around the same energy scale we expect the SM to break down. But already with the increasing popularity of models with large extra dimensions, also called Arkani-Hamed-Dimopoulos-Dvali models (ADD) \cite{Arkani-Hamed:1998jmv, Antoniadis:1998ig, Arkani-Hamed:1998sfv}, it became clear that even with a gravitational cutoff living in the UV this model is accompanied by a large number of additional degrees of freedom the so-called Kaluza-Klein (KK) modes. These new particles were by far much lighter than the cutoff of SM and it was noted that such theories have a different way of explaining neutrino masses without relying on heavy particles  \cite{Arkani-Hamed:1998wuz} like the seesaw mechanism but a large amount of additional light degrees of freedom suppresses the neutrino mass. This was by the knowledge of the author the first solution to the neutrino mass problem that relies on infrared physics without just accepting a very small Yukawa coupling. Theories that rely on additional light degrees of freedom to solve the neutrino mass problem can be called an infrared solution (IR solution). 

To date, two different classes of infrared theories exist. The first possibility is that neutrino masses are interlinked with the gravitational $\theta$-term \cite{Dvali:2016uhn}. The second class is the one we focus on in this work, under which the ADD model also falls. This class of theories uses a high number of mixing partners with the neutrino to suppress its mass. So far  
two models of that kind are known, the aforementioned ADD model and the Dvali-Redi (DR) model with many copies of the SM \cite{Dvali:2009ne}. These two models have in common that they solve the hierarchy problem by lowering the fundamental scale of gravity, $M_f$, down to TeV scale according to the formula \cite{Dvali:2007hz, Dvali:2007wp}
\begin{equation}
    M_f \leq \frac{M_P}{\sqrt{N}} \;,
    \label{speciesbound}
\end{equation}
in which $M_P$ is the Planck scale and $N$ is the number of species present in the theory. Additionally, in \cite{Ettengruber:2022pxf} it was shown that small neutrino masses are not just a model-specific property of the ADD and DR model but are a general feature of such theories. Therefore, solving the hierarchy problem by a low cutoff of gravity and the generation of small neutrino masses are intertwined in these theories. 

From a phenomenological point of view, these theories also offer an interesting way to test them in neutrino experiments as they lead to model-specific oscillation patterns like in ADD \cite{Arkani-Hamed:1998wuz,Dvali:1999cn} or the DR model \cite{Dvali:2009ne, Ettengruber:2022pxf} and therefore have been subject to plenty of experimental tests \cite{Machado:2011jt,Machado:2011kt,Basto-Gonzalez:2012nel,Girardi:2014gna,Rodejohann:2014eka,Berryman:2016szd,Carena:2017qhd,Stenico:2018jpl,Arguelles:2019xgp,DUNE:2020fgq,Basto-Gonzalez:2021aus,Arguelles:2022xxa} and \cite{Ettengruber:2024fcq}, respectively.

In this work, we want to extend the class with many additional mixing partners for the neutrino by another model which goes under the name of $N$naturalness \cite{Arkani-Hamed:2016rle}. Again this model was originally invented to solve the hierarchy problem but instead of solving it by lowering $M_f$ to the TeV scale, this model uses a cosmological selection scenario. 
Nevertheless, the mechanism to generate small neutrino mass stays the same as is shown in the following section \ref{Model}. Then a brief description of the ADD model is presented and similarities between ADD and $N$naturalness are discussed \ref{ADDintermezzo}. Afterward, the phenomenology for neutrino oscillations in experiments will be worked out in \ref{Phenomenology}, and in \ref{Discussion} the results of this work are discussed. Finally, in \ref{Conclusion} conclusions of the findings are drawn.

\section{$N$naturalness and Neutrino masses}
\label{Model}

The $N$naturalness scenario solves the hierarchy problem by arguing that the Higgs mass parameter $\mu^2$ of the SM is not a special choice but rather natural if one assumes a landscape of many possible $\mu_i^2$ of hidden higgsed sectors. These Higgs sectors are assumed to have a cutoff at the energy scale $\Lambda_H$ where either Higgs stabilizing physics like SUSY takes over or $\Lambda_H$ coincides with the fundamental scale of gravity $M_f$ that deviates from the Planck scale by $1/\sqrt{N_{sp}}$, where $N_{sp}$ is the number of additional species in the theory and can be quite large \cite{Dvali:2007hz, Dvali:2007wp}. That means that the hierarchy problem is solved by physics appearing at $\Lambda_H$ and $N$naturalness takes care of the splitting between the weak scale and $\Lambda_H$. 

The second ingredient for the mechanism is to assume the existence of number $N$ dark sectors that have a Higgs sector like the SM but one allows that the mass parameter $\mu_i^2$ can take every value between $[-\Lambda_H^2, \Lambda_H^2]$. This is the aforementioned landscape of higgs mass parameters $\mu_i$. If the additional $\mu_i$´s uniformly populate the interval $[-\Lambda_H^2, \Lambda_H^2]$ then the spacing among them will be in the precision of $-\frac{\Lambda_H^2}{N}$. 

The sectors with $\mu_i >0$ experience spontaneous symmetry breaking on a much lower scale and therefore the particles of such sectors are extremely light. Sectors with $\mu_i <0$ on the other hand will experience symmetry breaking like in the SM and the mass expressions of gauge bosons and fermions are similar to the SM and scale with the resulting vacuum expectation value (VEV) of the sector. The sector with the lightest VEV we identify as our SM. This explains why a small $\mu^2$ as observed is a natural value but does not explain why our SM is the one that was chosen by nature. In the original paper \cite{Arkani-Hamed:2016rle} the authors present cosmological models that show how in such a setup a cosmological history can be realized that dominantly populates our SM.

To let the statistical argument for the existence of a $\mu^2 = \frac{\Lambda_H^2}{N}$ hold, the authors furthermore argue that an SM structure in the Dark Sectors should not be atypical and the distribution of the $\mu_i$ should not show any preference within $[-\Lambda_H^2, \Lambda_H^2]$. Therefore, even if it is not a strictly necessary assumption it is natural to assume a uniform distribution among the $\mu_i$´s, and from this follows the expression
\begin{equation}
    \mu^2_i = -\frac{\Lambda^2_H}{N} (2i+r) \; ,
    \label{massparameter}
\end{equation}
where $i$ runs from $1,...,N/2$ and $r$ parametrizes the tuning of how much the $\mu^2$ values between our sector to the next heavier one deviate from the uniform splitting. This means that if $r=1$ the uniform distribution is realized and for $r<1$ the splitting is larger than one would expect from a $1/N$ precision. Translating equation \eqref{massparameter} into an expression for the VEV we get
\begin{equation}
    v_i = \Lambda_H \sqrt{\frac{2i}{\lambda N} + \frac{r}{\lambda N}} \; ,
    \label{veveq}
\end{equation}
where $\lambda$ is the usual quartic coupling of the SM Higgs sector. So the overall scale of of our VEV is $\Lambda_H/\sqrt{N}$. From this relation together with the lowering of $M_f$ according to \eqref{speciesbound} two scales for $\Lambda_H$ and $N$ are suggested. If one wants to keep the unification of the gauge couplings the number of present species should be lower than $N \leq 10^4$ which in turn would mean that $\Lambda_H \sim 10 \textrm{ TeV}$. The second suggestive choice of scales is when $M_f = \Lambda_H$ so when hierarchy problem is solved by lowering $M_f$ to the cutoff scale of the theory. The point where this happens is when $\Lambda_H = 10^{10} \textrm{ TeV}$ and $N = 10^{16}$.

To realize a Cosmology that makes the $N$naturalness scenario possible, in the original paper post inflationary models have been presented. Either a scenario with a fermionic reheaton $S$ with 
\begin{equation}
    \mathcal{L}_l \supset - \kappa S^c \sum_i l_i H_i - m_S SS^c \; ,
    \label{Sreheaton}
\end{equation}
or a with a scalar reheaton $\phi$ with
\begin{equation}
    \mathcal{L}_{\phi} = -a \phi \sum_i |H_i|^2 - \frac{1}{2}m_{\phi}^2 \phi^2\; .
\end{equation}
A crucial aspect of the cosmological models is that the masses of the reheatons should obey $m_{\textrm{reheaton}} \leq \Lambda_H/ \sqrt{N} $ which means $\mathcal{O}(m_{\textrm{reheaton}}) = 100 \textrm{ GeV}$.

From this point on the original discussion gets extended to the question of neutrino mass. To start with we discuss the two options of how neutrino masses arise in the $N$naturalness scenario and how they are implemented in the concrete models presented in \cite{Arkani-Hamed:2016rle}. 

The first option follows the line of argumentation of how neutrino masses are usually generated in IR models \cite{Arkani-Hamed:1998wuz, Dvali:2009ne, Ettengruber:2022pxf}. The crucial point is that the right handed neutrino $\nu_R$ is a singlet under the SM gauge group.

Therefore, if every sector is equipped with a right-handed neutrino the left-handed counterparts can interact with all other right-handed neutrinos even if they are not part of their own sector. In other words, the neutrino of our sector experiences mixing with $N$ right handed neutrinos. Of course, the left handed neutrinos of the other sectors will experience the same effect. This means you can create the following Dirac operator for neutrino masses
\begin{equation}
    (HL)_i \lambda_{ij}\nu_{Rj}\; ,
    \label{Diracoperator}
\end{equation}
in which the subscripts $i,j$ label the different sectors and the $\lambda_{ij}$ are Yukawa couplings constants which form a $N \times N$ matrix in the sector space. The SM sectors do not differ from another except of $\mu_i$ and therefore one would expect a permutation symmetry among them. This leads to the following structure
\begin{equation}
 \lambda_{ij}=\begin{pmatrix}
 a&b&b&\dots\\
 b&a&b&\dots\\
 b&b&a&\dots\\
 \dots&\dots&\dots&\ddots
\end{pmatrix}. 
\label{Yukawa}
\end{equation}
Note that the structure of this Matrix results from the coupling of different SM like sectors which was discussed in \cite{Dvali:2009ne}.
The off-diagonal elements have to obey the perturbative bound of 
\begin{equation}
    b \leq \frac{1}{\sqrt{N}} \; .
\end{equation}
Let us turn here to the relation between the two parameters $a$ and $b$. From the original construction of the $N$naturalness scenario where a soft permutation symmetry among the sectors is introduced that is just broken by the different values for the Higgs mass, the theory implies that the aforementioned Yukawa coupling should obey the rule of democratic mixing which would mean $a=b$. To take into account that the mixing among species of the same sector could slightly deviate from this general behaviour, we parametrize this by introducing $a$ but of course the natural expectation is $a \approx b \sim 1/\sqrt{N}$.

 The Dirac operator can in principle be present in both models but the S reheaton model also offers the possibility that a Weinberg operator is present in the effective theory of \eqref{Sreheaton} in the form of 
\begin{equation}
\frac{1}{m_S}(\bar{L}^ci\sigma_2 H)_i \lambda_{ij}(H i \sigma_2 L)_j \,.
\label{Weinberg}
\end{equation}
This scenario convolutes a Seesaw I scenario with the suppression of the coupling strength in IR scenarios. This means that the resulting mass of the neutrino is suppressed by the reheaton mass $m_S$ and by the fact that the coupling of this operator scales with $\lambda_{ij}\sim \kappa^2 \sim 1/N$ coming from \eqref{Sreheaton}. In the $N$naturalness scenario, the $1/N$  suppression in the Weinberg case is crucial because to let the cosmology work, a light reheaton of order $100 \text{GeV}$ is needed which is not enough to explain neutrino masses. 

Independently, which of the two operators is responsible for neutrino masses, both of them lead to similar neutrino mass matrices that just differ in their overall scaling. The diagonalization procedure is the same in both cases and is in detail explained in the Appendix so here just the final results are presented. 

The first finding is that in the case of a total democratic mixing, so $a =b$, no tower of heavier neutrino mass eigenstates exists. The matrix has $N-1$ degenerated eigenvalues of zero and one heavy eigenstate whose mass scales with $N$ or $N^2$ in the Dirac or Majorana case respectively. Of course, this option is already ruled out by neutrino oscillation measurements from where we know that the neutrino mass should be actually unequal to zero. 

As soon as one breaks the democratic coupling such that $a> b$ the degeneracy is lifted and a tower of additional neutrino mass eigenstates appears as one would naively expect from the increase of $v_i$.

In the Dirac case, the neutrino masses are 
\begin{equation}
    m_i = \sqrt{(a - b)^2\frac{\Lambda_H^2}{\lambda N}(2i+r + \mathcal{O}(\sin{\theta_i}^2))} \,,
    \label{Diracmass}
\end{equation}
and by taking $a - b \rightarrow \frac{1}{\sqrt{N}}$ we can simplify it to 
\begin{equation}
    m_i^D \approx \frac{1}{\sqrt{N}} v_i \,,
\end{equation}
as expected. From $N$naturalness two values for $N$ are suggested namely $N = 10^4$ and $N = 10^{16}$. Obviously, a suppression with $N= 10^{4}$ is not enough to explain neutrino masses without setting $a$ and $b$ to very similar values by hand. For $10^{16}$, the suppression reduces the required cancellation between $a$ and $b$ down to $10^{-3}$ which is less than the hierarchy among Yukawa couplings present within the SM of order $10^{-6}$. 

In the Majorana case, the mass goes as 
\begin{equation}
    m_i = \frac{v_i^2}{m_S}(a^2 - b^2)\,,
\end{equation}
which is for $(a^2 - b^2) \approx \frac{1}{N}$ in our sector
\begin{equation}
    m_{0} =\frac{v_0^2}{m_S N} \,.
    \label{MajoranaMass}
\end{equation}
For $N = 10^4$ the tuning is reduced already down to SM scales and for $N= 10^{16}$ the neutrino mass of our sector would be 
\begin{equation}
    m_0 \approx 6.05 \times 10^{-5} \textrm{ eV} \,.
\end{equation}

So without further requirement for taking Yukawas to small values by hand the $N$naturalness scenario predicts a mass for the neutrino that is in the ballpark as where one would expect it. The mass eigenstate composition of a neutrino of our sector looks as 
\begin{equation}
    \ket{\nu_0} = \ket{\nu_0}_{m} + \frac{1}{N} \sum_{i=1} \frac{\sqrt{2i+r}\sqrt{r}}{2i}\ket{\nu_i}_m + \frac{1}{N} \ket{\nu_N}_m\,,
    \label{massdecomposition}
\end{equation}
where we denoted the mass basis with the subscript $m$ and neglected an overall normalization factor. The mixing behavior can be understood qualitatively by recognizing that the one non-degenerated state in the democratic case exactly mixes with $1/N$. In the broken case, the mixing of the other states should then differ by 
\begin{equation}
    \frac{m_{ij}}{m_{ii}-m_{jj}}\,,
\end{equation}
where the $m_{ij}$ represent the elements in the mass matrix and can be read off leading to 
\begin{equation}
    \frac{\sqrt{2i+r} \sqrt{r}}{2i}\,.
\end{equation}

\section{A short ADD intermezzo}
\label{ADDintermezzo}
Before we proceed to the phenomenology of the $N$naturalness scenario it is helpful to have a look at the more familiar ADD model as both models show some similarities. For this reason, it is useful to recapitulate the findings of \cite{Arkani-Hamed:1998wuz, Dvali:1999cn} where the neutrino mixing with KK modes was analyzed.

KK modes appear if one compactifies additional spatial dimensions. In ADD one assumes that the extra dimensions are compactified on tori with radii of $R_i$. Furthermore, it is assumed that the SM particles live on the 3-dimensional subset of the 3+N dimensional world that is not compactified. Only particles that are not charged under the SM are allowed to propagate into the extra dimensions and can form a KK tower. The first candidate for such a particle is the graviton and the resulting KK tower exhibits $N=10^{32}$ KK modes if $R_i \approx \mu m$. With \eqref{speciesbound} the fundamental scale of gravity, $M_f$, gets lowered to the TeV scale and therefore this model offers a solution to the hierarchy problem. 

A second particle that is sterile under SM is the right-handed neutrino $\nu_R$. Through the mixing with the left-handed neutrino of the SM via a Dirac operator like \eqref{Diracoperator} it can also mix with the KK tower and the resulting mass decomposition of a neutrino in ADD can be written as 
\begin{equation}
    \nu_{\textrm{ADD}} = \nu_0^m + \xi \sum_{n=1}\frac{1}{n}\nu_n^m \;,
    \label{ADDtower}
\end{equation}
with
\begin{equation}
    \xi = \frac{\sqrt{2}hvM_fR}{M_P} \;,
\end{equation}
where $h$ is an order one coupling and these expressions have been calculated under the assumption that one radius, R, is much larger than the rest of them. The additional mass eigenstates $\nu_n^m$ have increasing masses with the mode number $n$ in the following way
\begin{equation}
    m_n = \frac{n}{R}\;.
    \label{ADDmasses}
\end{equation}

Already by comparing \eqref{ADDtower} with \eqref{massdecomposition} one sees the similarity of many additional mass modes contributing to the composition of the neutrino. So as in ADD a tower of light neutrino states appears in $N$naturalness but the source of it is not the compactification of spatial extra dimensions, it is the vacuum structure of the additional Higgs sectors.

The strength of mixing becomes weaker in both cases for the more massive modes but the scaling is different. The mixing of the KK-modes in the ADD case decreases with $\sim 1/n$ but the mixing in $N$naturalness just shows a $\sim 1/\sqrt{i}$ suppression. This means that the higher modes play a more important role than in ADD. Additionally, one has the heaviest mode that still mixes with $1/N$ which means that it mixes approximately with the same strength as the lightest one. 

If one looks at the masses one sees the reversed scaling. The ADD KK tower scales linearly with the mode number (see \eqref{ADDmasses}) meanwhile the $N$naturalness tower just has increasing masses of $\sqrt{2i+r}$ in the Dirac case (see \eqref{Diracmass}). 

This close relationship between ADD and $N$naturalness in the form of particle towers can be used as a guideline which phenomenological consequences could be of interest for testing $N$naturalness with terrestrial experiments. The physics of KK towers have been studied in depth and it is natural to ask how physics changes if one uses \eqref{massdecomposition} instead of \eqref{ADDtower}. The consequences for neutrino experiments are discussed in the following section \ref{Phenomenology} but also other signals are briefly discussed in \ref{Discussion}.
\section{Phenomenology}
\label{Phenomenology}
The mixing of the neutrinos with their counterparts across the different sectors has plenty of phenomenological consequences. To begin with, it is useful to calculate the survival probability of a neutrino of our sector
from \eqref{massdecomposition} which can be written as
\begin{equation}
    P_{surv} (L/E) = \left|1 + \frac{1}{N^2} \sum_{i=1} \frac{(2i+r)r}{4i^2} e^{i\phi_i} + \frac{1}{N^2} e^{i\phi_N}\right|^2 \,,
\end{equation}
with 
\begin{equation}
    \phi_i = |m_0^2 - m_i^2|\frac{L}{2E}\,,
\end{equation}
where $L$ is the distance of travel of the neutrinos from their production point and $E$ is their energy. If one averages over all modes with $n \geq2$ we get 
\begin{multline}
    P_{surv}(L/E)=  \left( 1+ \frac{(2+r)r}{4N^2}  \right)^2+\frac{1}{N^4} + \\\\- \frac{(2+r)r}{N^2} \sin^2\left({\frac{ (m_1^2 - m_0^2)  L}{4E}}\right) + 
    \frac{1}{N^4}\sum_{i=2}\left(\frac{(2i+r)r}{4i^2} \right)^2\; .
    \label{Eq:TOSC2}
\end{multline} 
Two important observations can be made from this analysis. The first is that two modes, the lightest and the heaviest, influence the neutrino oscillations with a strength of $1/N^2$. The second observation is that even though the tower of neutrino states is coming from a totally different source than a KK tower which one has in extra-dimensional theories, their phenomenology in neutrino oscillation experiments are somehow similar as we already argued in \ref{ADDintermezzo}. The point is that both theories predict a large number of states with decreasing mixing strength which allows us to average over the contribution of the heavier modes that we cannot resolve in experiments.

The two parameters that influence the mixing oscillation behaviour of neutrinos in this theory are $N$ and $r$. Both of them modulate the amplitude of the oscillation as can be seen from \eqref{massdecomposition} but do this in a sighlty different way. Meanwhile the $N$ sets the overall suppression and the hight of the tower, the $r$ controls the mixing especially for lower mass modes by the term $\sqrt{2i+r}/2i$. As explained in the previous section the role of $N$ is additionally to suppress neutrino masses with an overall scaling but this happens equally to all states and does therefore not influence their mass squared differences, $\Delta m_{ij}^2$, further. The influence of the $r$ parameter to the $\Delta m_{ij}^2$ can also be neglected in the first order in neutrino oscillations as the mass squared differences cancel the $r$ contribution which can be seen from expression \eqref{Diracmass}.

This leads us to a very special situation where the $\Delta m_{ij}^2$'s are predicted fully by the theory. We can calculate the mass differences with respect to the mass of our neutrino $\Delta m_{i0}^2 = m_{i}^2 - m_0^2$ by using the equations \eqref{Diracmass} and \eqref{MajoranaMass} for the Dirac and Majorana case, respectively. The scaling of 
\begin{equation}
 \Delta m_{i0}^2 \sim \frac{2\Lambda_H^2}{\lambda N^2} i   \textrm{\quad Dirac Case}\; ,
\end{equation}
is a smoking gun signature of the theory in the Dirac case and makes it distinguishable from other IR neutrino models. 

Also, the Majorana case exhibits such a smoking gun signature with a slightly different scaling of
\begin{equation}
    \Delta m_{i0}^2 \sim \frac{4 \Lambda_H^4}{m_S^2 \lambda^2 N^4}(i^2 + ir) \textrm{\quad Majorana Case}\; .
\end{equation}
This is also a special situation because the Majorana case and the Dirac case are actually distinguishable by neutrino oscillation experiments. The reason for this is that in the Majorana case, the masses scale with $v_i^2$ instead of $v_i$ as in the Dirac case. As $N$naturalness relies on a structure among the Higgs VEVs of the dark sectors the scaling of the neutrino masses gets affected accordingly.

The additional property of different scaling together with the fact that the scaling itself is a prediction of the model makes this model very interesting for experimental tests. 

For example could neutrino oscillation experiments search for the imprints of $N$naturalness in their neutrino spectra. In order to do so it is necessary to generalize \eqref{massdecomposition} to a realistic three flavor scenario. For a first approximation it is useful to assume a flavor symmetry in the mixing between the neutrinos of different sectors. If this is realized in the model, is for the phenomenology to first order less important as one expects cross-flavor mixing to be subdominant. This assumption leads to the following mass decomposition for a neutrino of flavor $\alpha$ of our sector as
\begin{equation}
    \ket{\nu_{\alpha 0}} = \sum_{j } \sum_{i=1}^3 U_{\alpha i} V_{0j} \ket{\nu_{ij}} \; ,
\end{equation}
where $i$ runs over the 3-flavor mixing matrix and $j$ runs over the additional sectors. The $U$ matrix is the well known lepton mixing matrix and $V$ represents the mixing matrix of the different sectors. 

\begin{figure}
        \centering
        \includegraphics[scale= 0.4]{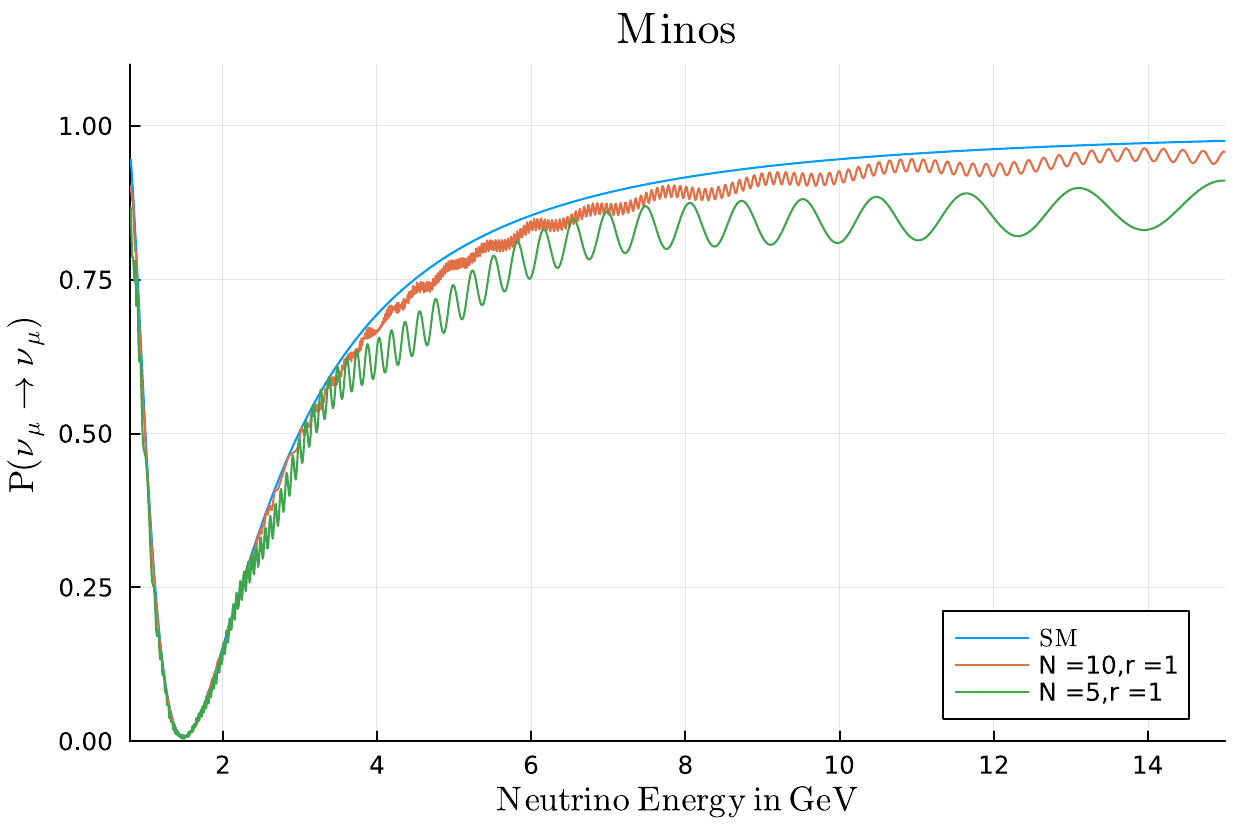}
        \caption{The survival probability of a $\nu_{\mu}$ in the Minos Far Detector as a function of the neutrino energy in the Dirac case. For SM oscillation parameters given by \cite{ParticleDataGroup:2024cfk}, normal ordering, and $m_0 = 0.01 \textrm{ eV}$. The blue line represents the ordinary three-flavor oscillation meanwhile the red and green lines represent the NNaturalness with $N=10, r=1$ and $N=5, r=1$, respectively.}
        \label{Minossurv}
    \end{figure}

\begin{figure}
        \centering
        \includegraphics[scale= 0.4]{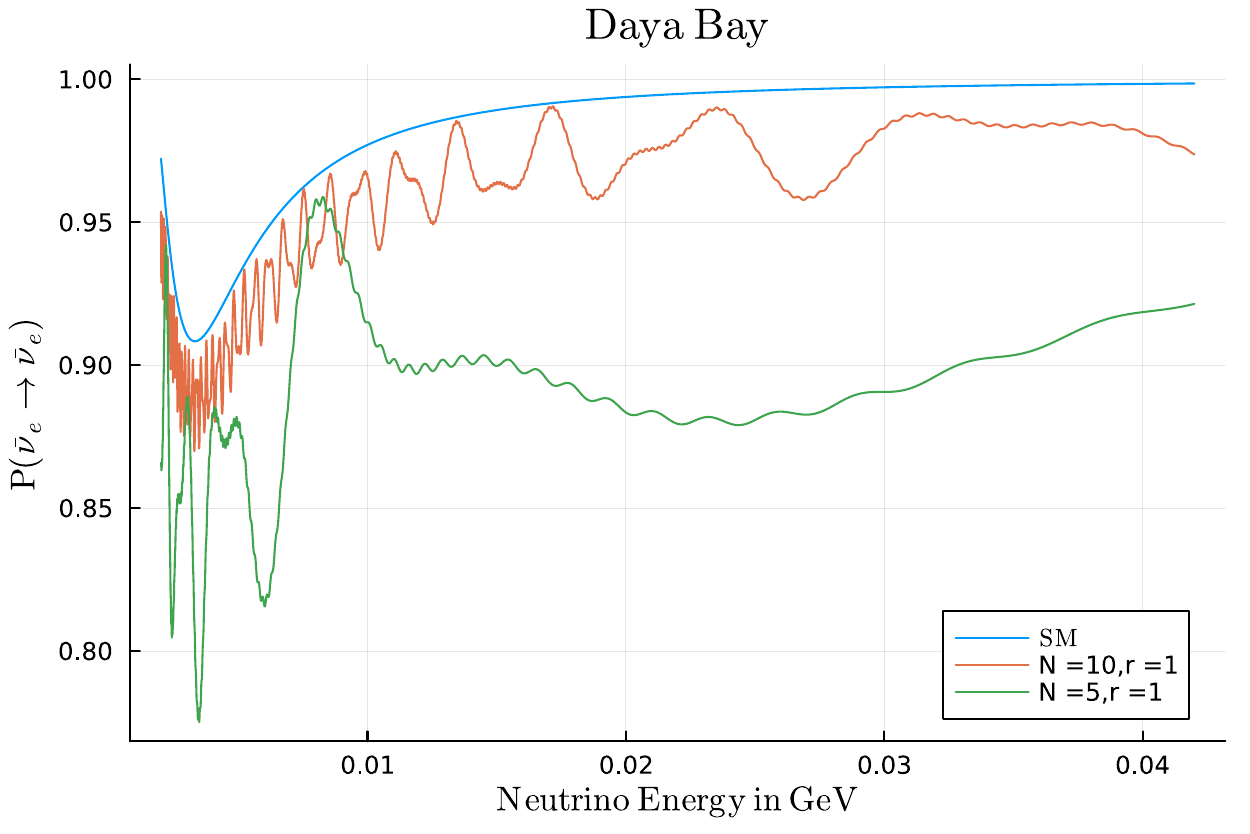}
        \caption{The survival probability of a $\bar{\nu}_e$ in the Daya Bay experiment in experimental hall 3 in the Dirac case. For SM oscillation parameters given by \cite{ParticleDataGroup:2024cfk}, normal ordering, and $m_0 = 0.01 \textrm{ eV}$. The blue line represents the ordinary three-flavor oscillation meanwhile the red and green lines represent the $N$naturalness with $N=10, r=1$ and $N=5, r=1$, respectively.}
        \label{Dayasurv}
    \end{figure}

Equipped with this equation one can investigate the oscillation behavior of neutrinos in experiments like Minos \cite{Evans:2013pka} and DayaBay \cite{Cao:2016vwh} and the results can be studied in the Fig. \ref{Minossurv} and Fig. \ref{Dayasurv}. As expected one sees that the oscillation amplitude decreases with increasing $N$ and converges to the SM for large N. The second observation one can make especially in \ref{Minossurv} is that two modes, the lightest and and the heaviest, influence the oscillation behaviour with approximately equal strength but their oscillation frequency is different. 

\begin{figure}
        \centering
        \includegraphics[scale= 0.4]{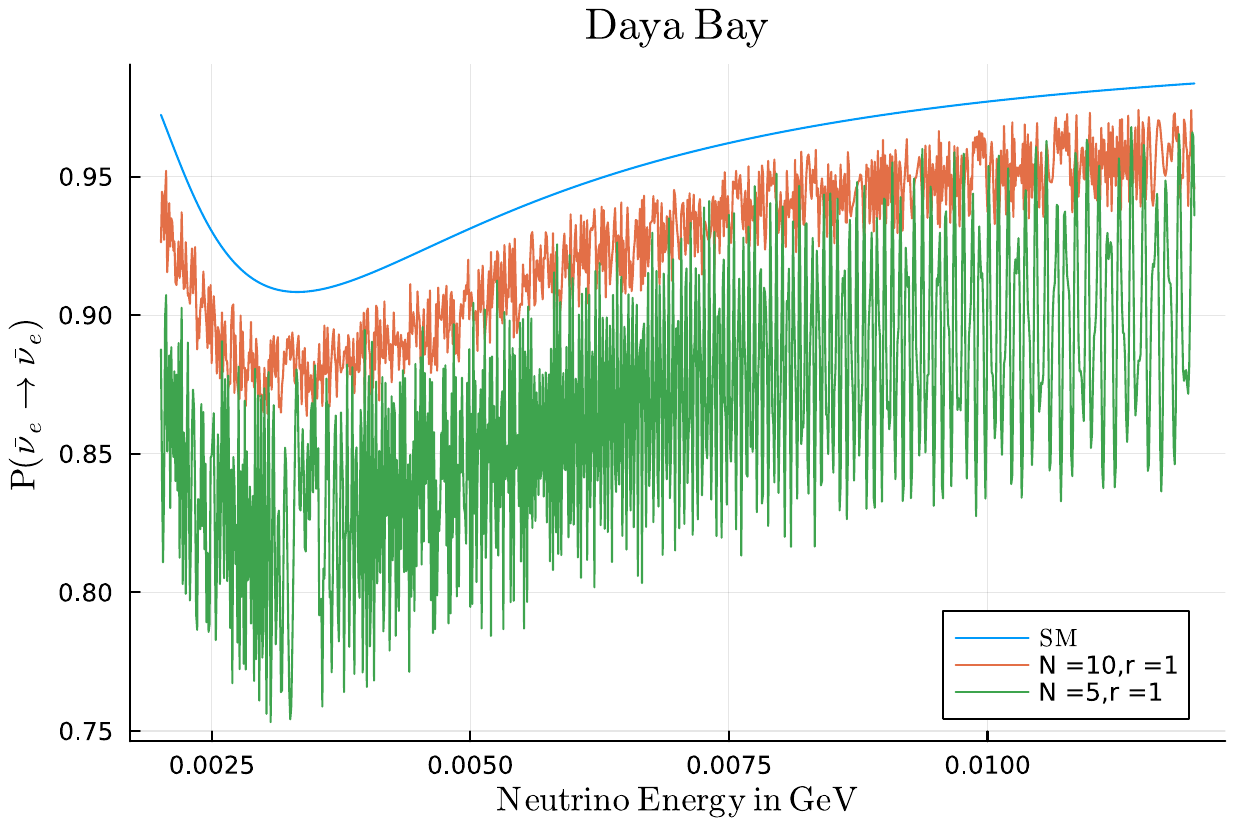}
        \caption{The survival probability of a $\bar{\nu}_e$ in the Daya Bay experiment in experimental hall 3 in the Majorana case. For SM oscillation parameters given by \cite{ParticleDataGroup:2024cfk}, normal ordering, and $m_0 = 0.01 \textrm{ eV}$. The blue line represents the ordinary three-flavor oscillation meanwhile the red and green lines represent the $N$naturalness with $N=10, r=1$ and $N=5, r=1$, respectively.}
        \label{DayasurvMajorana}
    \end{figure}

In Fig. \ref{DayasurvMajorana} the oscillation behavior in the DayaBay Experiment is shown for the Majorana case and can be compared with Fig. \ref{Dayasurv} which shows the Dirac case. A clear difference in the oscillation frequency can be seen and that is the observable that allows us to discriminate between the Dirac and Majorana case.

For this scenario like in ADD matter effects could play a crucial role for detection because the existence of many additional modes makes it easier to fulfill the conditions for a resonant enhancement of the signal. In general, these effects can be calculated with the following expression for the effective Hamiltonian

\begin{multline}
H_{eff}  = \frac{1}{2E} \Biggl[ U 
\begin{pmatrix}
m_e & 0& 0 & 0 &0\\
0 &m_{\mu}&0 &0&0\\
0 & 0 &m_{\tau}&0&0\\
0&0&0&m_4&0\\
0&0&0&0&\ddots
\end{pmatrix}U^{\dagger}  \\\\
+ \begin{pmatrix}
A & 0& 0 & 0 &0\\
0 &0&0 &0&0\\
0 & 0 &0&0&0\\
0&0&0&A^{\prime}&0\\
0&0&0&0&\ddots
\end{pmatrix}\Biggr] \;,
\end{multline}
with $A = 2 \sqrt{2} G_F N_e E$, $A^\prime = - \sqrt{2}G_F N_n E$ and $N_e$, $N_n$ being the densities of electrons, neutrons respectively. The resonance condition resulting from this effective Hamiltonian reads as
\begin{equation}
    V = \frac{\Delta m_{0i}^2}{2E}\cos{2\theta} \approx\frac{\Delta m_{0i}^2}{2E} \;,
\end{equation}
where $V$ is 
\begin{equation}
    V = \sqrt{2} G_F \left(N_e - \frac{N_n}{2}\right) \;.
\end{equation}
The enhancement by matter effects is particularly interesting for experiments that are already measuring such effects in the usual three-flavor massive neutrino paradigm like IceCube \cite{Ahlers2018}.

    Another type of experiment that will experience influence from the many additional light states are neutrino mass measurement experiments like KATRIN. The higher massive modes can be in principle resolved by the KATRIN experiment and could therefore be used to constrain the parameters of the theory. In order to do so a full shape analysis of the KATRIN spectrum would be necessary like in the ADD case. The striking similarities between $N$naturalness and ADD allows us to just modify the Kurie function for ADD \cite{Basto-Gonzalez:2012nel} with our theory parameters and get
    \begin{multline}
        K(E, m_0, N, r) = \sum_{k}p_k \epsilon_k \sum_{i}|U_{\alpha i}|^2 \sum_{j} |V_{0 j}|^2  \\\\\sqrt{\epsilon_k^2 - m_{i,j}^2} \theta(\epsilon_k - m_{i,j}) \; ,
    \end{multline}
where $\epsilon_k = E_0 - V_k - E$ is the neutrino energy which depends on the endpoint of the spectrum $E_0$ and $V_k$ are the different energies of the final states that occur with the probability $p_k$.  Again like in the neutrino oscillation case the different scaling in the neutrino masses of higher modes allows such experiments to discriminate between the Majorana and Dirac case. 

This brings us to the third type of low energy particle experiments namely neutrinoless double beta decay (0$\nu\beta\beta$) searches. The crucial parameter rooted in fundamental properties of the neutrino mass is the so called effective Majorana mass, $m_{\beta \beta}$ which is in the three-flavor case
\begin{equation}
    m_{\beta \beta} = \sum_i U_{ei}^2 m_i \; .
    \label{effektivemajoranamass}
\end{equation}
In the light neutrino case, this quantity is the only one influenced by neutrino masses meanwhile the nuclear matrix elements (NME) that govern the decay behavior of atoms that could perform 0$\nu\beta\beta$ decays are independent of them. This is not necessarily true for heavy neutrino mass modes which we would expect from $N$naturalness. Therefore, one can say that in principle these experiments are suited to constrain Majorana scenarios of $N$naturalness. The influence of it is maybe not as trivial as just adjusting \eqref{effektivemajoranamass} and to analyze this case goes beyond of the scope of this paper. 

\section{Discussion}
\label{Discussion}
In the previous section has been worked out how signatures of a $N$naturalness scenario can show up in the properties of neutrinos. So far the research for such scenarios was focused on their cosmological consequences. This is also true for work that considers the neutrino sector \cite{Han:2018pek, Bansal:2024afn}. The result of this paper is that low energy neutrino experiments can be used to give lower bounds on the number of additional neutrino mixing partners. The next generation of oscillation experiments like JUNO \cite{JUNO:2015sjr} and DUNE \cite{DUNE:2020lwj} that will decrease the uncertainty in the lepton mixing parameters by an order of magnitude offer the exciting possibility to perform an actual terrestrial test of the $N$Naturalness scenario. 

Of course, by knowing the exact neutrino mass composition and therefore the mixing and oscillation behavior for such scenarios, helps cosmological and astrophysical research as it allows a dedicated test of $N$naturalness. For example, could a similar analysis of neutrino cosmology like in the case of the DR model \cite{Zander:2023jcu} be used to either restrict the model parameters or give bounds on the maximal possible reheating temperature.

But not just neutrinos could experience a mixing with their heavier counterparts. Another candidate for this would be the neutron as it is also not charged under the SM. This is not surprising as something similar has already been suggested in the DR model \cite{Dvali:2009ne} as well as in the ADD model \cite{Dvali:1999gf, Dvali:2023zww}. In \cite{Dvali:2023zww} it was particularly shown in the case of the neutron mixing with a KK tower that the phenomenological consequences can be quite severe. The reason for this is that the mixing of bounded neutrons with dark partners has to be small enough to not violate neutron lifetime bounds of order $10^{30}$ yrs \cite{KamLAND:2005pen}. Something similar could happen in $N$naturalness but such an analysis goes again beyond this paper. Therefore neutron experiments could also play an important role in testing $N$naturalness. 

In all aforementioned experimental cases for detection, it would be of advantage if $N$ is not too large. During the discussion of neutrino masses it was argued that in order to solve the neutrino mass problem, a scenario with just $N = 10^4$ would not be sufficient to explain the smallness of neutrino masses if one whishes no special choice for Yukawas at all. This requirement is maybe to restrictive as we already see a hierarchy within the SM of order $10^{-6}$. In the case of $N=10^{16}$ no special choice for small differences among the Yukawas is required but one has to keep in mind that scenarios with such a large number of additional Yang-Mills sectors (in case the QCD sector is also part of the dark sectors) may suffer from consistency requirements of quantum gravity with axion physics \cite{Ettengruber:2023tac}. On the other hand if $N$ is not too large like $N=10^4$ then the situation for axions changes and such scenarios can alter the coupling behavior of the axion \cite{Ettengruber:2023tac} which is an additional potential signal of the DR and the $N$naturalness model. 

However, from a phenomenological point of view, it is not necessary that the overall suppression of the mass matrix only comes from the mixing with many partners. A small Yukawa coupling is still possible or other physics is taking care of the overall suppression of neutrino masses. Nevertheless, the described operators would still appear and the $N$naturalness phenomenology would still show up in experiments. 

Another aspect that is worth mentioning is that with extra-dimensional theories and the DR model the $N$naturalness scenario is the third class of theories that exhibit a suppression of the neutrino mass due to the existence of many mixing partners. Therefore, one is tempted to say that this way of neutrino mass generation is a general mechanism that can be implemented in specific models. Like in seesaw scenarios, the specific phenomenology depends on the concrete model but the general mechanism stays the same. Compared to UV models where heavy particles are responsible for the small mass of the neutrino rely the IR models on the existence of many additional light mixing partners for the neutrino.

\section{Conclusion}
\label{Conclusion}
In conclusion this work has shown that $N$naturalness is able to solve the neutrino mass problem. In addition the neutrino phenomenology of $N$naturalness could be much more rich than originally anticipated. Either the necessary ingredients have already been part of the original proposals like in the fermionic reheaton model where a Weinberg operator already exists on tree level, or just need small additional extensions like right handed neutrinos that can form a Dirac operator with the left handed SM ones. Both cases lead to the same mixing beviour of neutrinos with their dark counterparts. 

Because the neutrino masses depend on the VEVs of the additional sectors the scaling of neutrino masses in $N$naturalness shows a distinctive pattern. An interesting effect of this is that the scaling is different between the Dirac and Majorana mass case which would then lead to different oscillation frequencies in neutrino oscillation experiments. Therefore, could such experiments actually discriminate between a Majorana and Dirac mass which is usually not the case in other neutrino mass models. 

The maybe most exciting outcome of this work is that low energy experiments like neutrino experiments could actually be suited to investigate such scenarios. Equipped with the equations of this analysis we can perform terrestrial experiments on $N$naturalness and bring the question of how to test $N$naturalness down from the sky to the earth. 

\section{Acknowledgements}
I am thankful to Raffaele Tito D´Agnolo for useful discussions about $N$naturalness and for giving helpful comments about the draft. This work was supported by ANR grant ANR-23-CE31-0024
EUHiggs.

\newpage
\section{Appendix}
\begin{widetext}
The mass matrix resulting from the Dirac operator \eqref{Diracoperator} is

\begin{equation}
M_{\text{Dirac}}= \begin{pmatrix}
a\sqrt{\frac{\Lambda^2}{\lambda N}}\sqrt{r} & b\sqrt{\frac{\Lambda^2}{\lambda N}}\sqrt{r} & b\sqrt{\frac{\Lambda^2}{\lambda N}}\sqrt{r}&... &b\sqrt{\frac{\Lambda^2}{\lambda N}}\sqrt{r}
\\ b\sqrt{\frac{\Lambda^2}{\lambda N}}\sqrt{2+r} & a\sqrt{\frac{\Lambda^2}{\lambda N}}\sqrt{2+r}& b\sqrt{\frac{\Lambda^2}{\lambda N}}\sqrt{2+r}&...&b\sqrt{\frac{\Lambda^2}{\lambda N}}\sqrt{2+r}\\

\vdots&&\ddots&&\vdots&\\

b\sqrt{\frac{\Lambda^2}{\lambda N}}\sqrt{2(N-1)+r}&&...&&a\sqrt{\frac{\Lambda^2}{\lambda N}}\sqrt{2(N-1)+r}\\

\end{pmatrix}\;.
\label{Dirmamat}
\end{equation}
The mixing of neutrinos can be calculated by diagonalizing 
\begin{equation}
    M^2 = M_{\text{Dirac}}M_{\text{Dirac}}^{\dagger}
\end{equation}
This is 
\begin{equation}
    M^2 =  \begin{pmatrix}
\Tilde{a}^2 r & \Tilde{b}^2 \sqrt{2+r} \sqrt{r} & \Tilde{b}^2 \sqrt{4+r}\sqrt{r}&... &\Tilde{b}^2\sqrt{2(N-1)+r}\sqrt{r}
\\ \Tilde{b}^2\sqrt{2+r}\sqrt{r} & \Tilde{a}^2 (2+r)& \Tilde{b}^2 \sqrt{4+r} \sqrt{2+r}&...&\Tilde{b}^2 \sqrt{2(N-1)+r} \sqrt{2+r}\\

\vdots&&\ddots&&\vdots&\\

\Tilde{b}^2 \sqrt{r} \sqrt{2(N-1)+r}&&...&&\Tilde{a}^2 (2(N-1)+r)\\

\end{pmatrix} \frac{\Lambda^2}{\lambda N}\;.
\label{masssqr}
\end{equation}
with 
\begin{equation}
    \Tilde{a}^2 = (N-1)b^2 + a^2,
    \label{atilde}
\end{equation}
\begin{equation}
    \Tilde{b}^2 = (N-2)b^2 + 2ba.
    \label{btilde}
\end{equation}
Similarly one can use the Weinberg operator which leads to the same matrix as in \eqref{masssqr}. 
To start the diagonalization procedure one first has to look into the special case of $\Tilde{a}=\Tilde{b}$ and for simplicity we drop the overall factor $\frac{\Lambda^2}{\lambda N}$. Then the first step is
\begin{equation}
    M^\prime = U_1^{\dagger} M^2 U_1
\end{equation}
with 
\begin{equation}
    U_1 = 
 \begin{pmatrix}
\cos{\theta} & \sin{\theta} & 0&\dots&\dots&0
\\ -\sin{\theta} & \cos{\theta}& 0&\dots&\dots&0\\
0&0&1&0&\dots&0\\
\vdots&&&\ddots&&\vdots&\\

0&&...&&&1\\

\end{pmatrix}
\end{equation}
where the mixing angle is given by 
\begin{equation}
    \theta = \frac{1}{2}\arctan{\sqrt{r}\sqrt{r+2}}
\end{equation}
After this operation, we arrive at
\begin{equation}
    M^\prime = \begin{pmatrix}
0  &\dots& \dots& \dots&\dots &0
\\ \vdots & 2(1 + r)& \sqrt{2} \sqrt{1 + r} \sqrt{4 + r}&\dots&\dots&\sqrt{2}\sqrt{2(N-1)+r} \sqrt{1+r}\\
\vdots& \sqrt{2}\sqrt{1 + r} \sqrt{4 + r}& 4+r& \sqrt{4+r} \sqrt{6+r}&\dots&\sqrt{2(N-1)+r}\sqrt{2+r} \\
\vdots&\vdots&&\ddots&&\vdots&\\
\vdots&\vdots&&&\ddots&\vdots&\\
0&\sqrt{2}\sqrt{2(N-1)+r} \sqrt{1+r}&...&&&2(N-1)+r\\

\end{pmatrix}
\end{equation}
From $M^\prime$ we can repeat the steps similar to the first one presented. By induction, one can show that the mixing angles behave as 
\begin{equation}
    \theta_i = \frac{1}{2} \arctan{\frac{2 \sqrt{i(i-1+r)}\sqrt{2i+r}}{2i+r - i(i-1+r)}} \; ,
\end{equation}
where $i$ stands for the number of the step. The resulting diagonalized mass matrix is then
\begin{equation}
        M_{\text{diag}}^2 = 
 \begin{pmatrix}
0 & 0 & 0&\dots&\dots&0
\\ 0 & 0& 0&\dots&\dots&0\\
0&0&N(N-1+r)&0&\dots&0\\
0 &0&0&0&\dots&0\\
\vdots&\vdots&\vdots&\vdots&\ddots&\vdots&\\

0&&...&&&0\\
\end{pmatrix} \; .
\end{equation}
This means that $N-1$ states are degenerated with a mass of zero and one state has a mass of $m_H = \sqrt{N(N-1+r)}$. The corresponding eigenstates are the columns of the diagonalization matrix
\begin{equation}
    V = U_1 U_2 ... U_N \; .
\end{equation}
From this diagonalization matrix, $V$ one can read off the structure of the degenerated eigenstates which is
\begin{equation}
    v_1 = \begin{pmatrix}
        \cos{\theta_1} \\
        -\sin{\theta_1}\\
        0\\
        \vdots
    \end{pmatrix}, v_2 = \begin{pmatrix}
         \sin{\theta_1} \cos{\theta_2} \\
        \cos{\theta_1} \cos{\theta_2}\\
        -\sin{\theta_2}\\
        0\\
        \vdots
    \end{pmatrix}, 
    v_3 = \begin{pmatrix}
        \sin{\theta_1}\sin{\theta_2} \sin{\theta_3} \\
        \cos{\theta_1} \sin{\theta_2} \sin{\theta_3}\\
        \cos{\theta_2}\sin{\theta_3}\\
        \cos{\theta_3}\\
        0\\
        \vdots
    \end{pmatrix}, 
        v_4 = \begin{pmatrix}
        \sin{\theta_1}\sin{\theta_2} \cos{\theta_3}\sin{\theta_4} \\
        \cos{\theta_1} \sin{\theta_2} \cos{\theta_3} \sin{\theta_4}\\
        \cos{\theta_2}\cos{\theta_3} \sin{\theta_4}\\
        -\sin{\theta_3} \sin{\theta_4}\\
        \cos{\theta_4}\\
        0\\
        \vdots
    \end{pmatrix}, v_5 = ... \; ,
    \label{masseigenstates}
\end{equation}
and the heavy eigenstates is 
\begin{equation}
    v_H = \begin{pmatrix}
        \sin{\theta_1} \sin{\theta_2} \cos{\theta_3} \dots \cos{\theta_N}\\
        \cos{\theta_1} \sin{\theta_2} \cos{\theta_3} \dots \cos{\theta_N}\\
        \cos{\theta_2} \dots \cos{\theta_N}\\
        -\sin{\theta_3}\cos{\theta_4} \dots \cos{\theta_N}\\
        -\sin{\theta_4} \cos{\theta_5}\dots \cos{\theta_N}\\
        \vdots\\
        -\sin{\theta_N}
    \end{pmatrix} \; .
\end{equation}
The first entry of $\nu_H$ always results in $1/N$. The second layer of the analysis is to deviate from the special case $\Tilde{a}=\Tilde{b}$ and treat the deviation from this case as a perturbation. Because both couplings are of the same nature one expects that their difference is of order $1/N$. The Ansatz is
\begin{equation}
    M^2_P = M^2 + P\; ,
\end{equation}
where 
\begin{equation}
    P = 
     \begin{pmatrix}
(\Tilde{a}^2-\Tilde{b}^2)r & 0 & 0&\dots&0
\\ 0 & (\Tilde{a}^2-\Tilde{b}^2)(2+r)& 0&\dots&0\\

0 &0&\ddots&\dots&0\\
\vdots&\vdots&\vdots&\ddots&\vdots&\\

0&&...&&(\Tilde{a}^2-\Tilde{b}^2)(2(N-1)+r)\\
\end{pmatrix} \; .
\end{equation}
Using degenerated perturbation theory one has to calculate 
\begin{equation}
    W_{ab} = \bra{\Psi^0_a}P\ket{\Psi^0_b} \; ,
\end{equation}
where the subscripts $a,b$ run over the number of degenerated states $\Psi^0_a$. This matrix can be calculated exactly and in order to find the "good" basis to perform perturbation theory one has to diagonalize $W$. Then the eigenvectors of this matrix tell how the good eigenstates are composed out of the original eigenvectors that are the columns of $V$. Because the off-diagonal elements of $W$ are suppressed by orders of magnitude in $\theta_i$´s compared to the diagonal elements, the $W$ can be perturbatively diagonalized. This procedure can achieve in principle any precision that is necessary for the calculation and the phenomenological prediction. The good eigenstates then just are
\begin{equation}
    \Psi^G_{\xi} = \alpha_{1\xi} \Psi_1 +  \alpha_{2\xi} \Psi_2 + \dots +\Psi_{\xi}+\dots+ \alpha_{N\xi} \Psi_N \; .
\end{equation}
The mixing of the original basis $\ket{\Psi}$ with the good basis $\ket{\Psi^G}$ can be expressed as
\begin{equation}
    \ket{\Psi^G} = \begin{pmatrix}
        1&\alpha_{12}&\dots&\dots&\dots&\alpha_{1N}\\
        -\alpha_{12} & 1 &\alpha_{23}&\dots&\dots&\alpha_{2N}\\
        \vdots&-\alpha_{23}&1&\alpha_{34}&\dots&\alpha_{3N}\\
        \vdots&\vdots&-\alpha_{34}&\ddots&\dots&\vdots\\
        
        \vdots&\vdots&\vdots&\vdots&\ddots&\vdots\\
        -\alpha_{1N}&\dots&\dots&\dots&\dots&1
    \end{pmatrix} \ket{\Psi} \; .
    \label{goodvsbadmatrix}
\end{equation}

The link between the sector basis and the good basis is then
\begin{equation}
    \ket{\Psi^G} = V U_P\ket{n} \; ,
\end{equation}
where $U_P$ is the matrix of \eqref{goodvsbadmatrix}. From this expression, one can see how strong the neutrino of the highest sector number is influencing the first mass eigenstate. The mixing angle goes to zero and therefore the last entry of $v_{N-1}$ goes to one. All other original eigenvectors have zeros in the last entry and therefore the contribution of $v_{N-1}$ to $\Psi_1^G$ is the only one and is 
\begin{equation}
    \alpha_{1N} \approx \frac{1}{N \sqrt{N}}\; .
\end{equation}
To do the summation over all original eigenvectors and weight them accordingly with their corresponding $\alpha_{ij}$´s is impossible but from the original mass basis, we can read off how the mixing among them behaves. The mixing contributing to the first mass eigenstate can be then described as
\begin{equation}
    \ket{\nu_0}_m = 
    \ket{\nu_0} + \frac{1}{N} \sum_{i=1} \frac{\sqrt{2i+r}\sqrt{r}}{2i}\ket{\nu_i}\,.
    \label{firstmasseigenstate}
\end{equation}
In order to find the expression for the first sector neutrino in the mass basis one has to keep in mind that there was one additional heavy mass eigenstate in the unperturbed case that mixed with the first sector neutrino with the strength of $1/N$. In principle, to finish the perturbative diagonalization procedure one would have to calculate the perturbation towards the $\ket{\Psi^G}$ basis rooting from this state. Nevertheless, for large $N$ the contribution to the lightest states can be neglected and we can give the final expression as
\begin{equation}
    \ket{\nu_0} = \ket{\nu_0}_{m} + \frac{1}{N} \sum_{i=1} \frac{\sqrt{2i+r}\sqrt{r}}{2i}\ket{\nu_i}_m + \frac{1}{N} \ket{\nu_H}_m\,,
\end{equation}
where the heavy mass eigenstate has to be reintroduced in the composition.  
\end{widetext}
\setlength{\bibsep}{5pt}

\bibliographystyle{utphys}
\bibliography{refs}

@article{Evans:2013pka,
    author = "Evans, Justin",
    collaboration = "MINOS",
    title = "{The MINOS Experiment: Results and Prospects}",
    eprint = "1307.0721",
    archivePrefix = "arXiv",
    primaryClass = "hep-ex",
    reportNumber = "FERMILAB-PUB-13-279-PPD",
    doi = "10.1155/2013/182537",
    journal = "Adv. High Energy Phys.",
    volume = "2013",
    pages = "182537",
    year = "2013"
}

@article{Cao:2016vwh,
    author = "Cao, Jun and Luk, Kam-Biu",
    title = "{An overview of the Daya Bay reactor neutrino experiment}",
    eprint = "1605.01502",
    archivePrefix = "arXiv",
    primaryClass = "hep-ex",
    doi = "10.1016/j.nuclphysb.2016.04.034",
    journal = "Nucl. Phys. B",
    volume = "908",
    pages = "62--73",
    year = "2016"
}

@Article{Ahlers2018,
  author   = {Ahlers, Markus and Helbing, Klaus and Pérez de los Heros, Carlos},
  journal  = {The European Physical Journal C},
  title    = {Probing particle physics with IceCube},
  year     = {2018},
  issn     = {1434-6052},
  number   = {11},
  pages    = {924},
  volume   = {78},
  refid    = {Ahlers2018},
  url      = {https://doi.org/10.1140/epjc/s10052-018-6369-9},
  note 		=  {arXiv:1806.05696v1 [astro-ph.HE], doi: 10.1140/epjc/s10052-018-6369-9}
}

@article{Han:2018pek,
    author = "Han, Zhi-Long and Zhu, Bin and Bian, Ligong and Ding, Ran",
    title = "{Consistent origin of neutrino mass and freeze-in dark matter in large N theories}",
    eprint = "1812.00637",
    archivePrefix = "arXiv",
    primaryClass = "hep-ph",
    month = "12",
    year = "2018"
}

@article{Bansal:2024afn,
    author = "Bansal, Saurabh and Ghosh, Subhajit and Low, Matthew and Tsai, Yuhsin",
    title = "{A cosmological case study of a tower of warm dark matter states: $N$naturalness}",
    eprint = "2410.19224",
    archivePrefix = "arXiv",
    primaryClass = "astro-ph.CO",
    reportNumber = "UTWI-31-2024",
    month = "10",
    year = "2024"
}

@article{Zander:2023jcu,
    author = "Zander, Alan and Ettengruber, Manuel and Eller, Philipp",
    title = "{How many dark neutrino sectors does cosmology allow?}",
    eprint = "2308.00798",
    archivePrefix = "arXiv",
    primaryClass = "hep-ph",
    doi = "10.1140/epjc/s10052-024-12689-7",
    journal = "Eur. Phys. J. C",
    volume = "84",
    number = "3",
    pages = "331",
    year = "2024"
}

@article{Dvali:1999gf,
    author = "Dvali, G. R. and Gabadadze, Gregory",
    title = "{Nonconservation of global charges in the brane universe and baryogenesis}",
    eprint = "hep-ph/9904221",
    archivePrefix = "arXiv",
    reportNumber = "NYU-TH-99-3-02",
    doi = "10.1016/S0370-2693(99)00766-2",
    journal = "Phys. Lett. B",
    volume = "460",
    pages = "47--57",
    year = "1999"
}

@article{KamLAND:2005pen,
    author = "Araki, T. and others",
    collaboration = "KamLAND",
    title = "{Search for the invisible decay of neutrons with KamLAND}",
    eprint = "hep-ex/0512059",
    archivePrefix = "arXiv",
    doi = "10.1103/PhysRevLett.96.101802",
    journal = "Phys. Rev. Lett.",
    volume = "96",
    pages = "101802",
    year = "2006"
}

@article{Ettengruber:2023tac,
    author = "Ettengruber, Manuel and Koutsangelas, Emmanouil",
    title = "{Consequences of multiple axions in theories with dark Yang-Mills groups}",
    eprint = "2307.10298",
    archivePrefix = "arXiv",
    primaryClass = "hep-ph",
    doi = "10.1103/PhysRevD.111.036006",
    journal = "Phys. Rev. D",
    volume = "111",
    number = "3",
    pages = "036006",
    year = "2025"
}

@article{DUNE:2020lwj,
    author = "Abi, Babak and others",
    collaboration = "DUNE",
    title = "{Deep Underground Neutrino Experiment (DUNE), Far Detector Technical Design Report, Volume I Introduction to DUNE}",
    eprint = "2002.02967",
    archivePrefix = "arXiv",
    primaryClass = "physics.ins-det",
    reportNumber = "FERMILAB-PUB-20-024-ND, FERMILAB-DESIGN-2020-01",
    doi = "10.1088/1748-0221/15/08/T08008",
    journal = "JINST",
    volume = "15",
    number = "08",
    pages = "T08008",
    year = "2020"
}

@article{JUNO:2015sjr,
    author = "Djurcic, Zelimir and others",
    collaboration = "JUNO",
    title = "{JUNO Conceptual Design Report}",
    eprint = "1508.07166",
    archivePrefix = "arXiv",
    primaryClass = "physics.ins-det",
    month = "8",
    year = "2015"
}

@article{Dvali:2023zww,
    author = "Dvali, Gia and Ettengruber, Manuel and Stuhlfauth, Anja",
    title = "{Kaluza-Klein spectroscopy from neutron oscillations into hidden dimensions}",
    eprint = "2312.13278",
    archivePrefix = "arXiv",
    primaryClass = "hep-ph",
    doi = "10.1103/PhysRevD.109.055046",
    journal = "Phys. Rev. D",
    volume = "109",
    number = "5",
    pages = "055046",
    year = "2024"
}

@article{Basto-Gonzalez:2012nel,
    author = "Basto-Gonzalez, Victor S. and Esmaili, Arman and Peres, Orlando L. G.",
    title = "{Kinematical Test of Large Extra Dimension in Beta Decay Experiments}",
    eprint = "1205.6212",
    archivePrefix = "arXiv",
    primaryClass = "hep-ph",
    doi = "10.1016/j.physletb.2012.11.048",
    journal = "Phys. Lett. B",
    volume = "718",
    pages = "1020--1023",
    year = "2013"
}

@article{ParticleDataGroup:2024cfk,
    author = "Navas, S. and others",
    collaboration = "Particle Data Group",
    title = "{Review of particle physics}",
    doi = "10.1103/PhysRevD.110.030001",
    journal = "Phys. Rev. D",
    volume = "110",
    number = "3",
    pages = "030001",
    year = "2024"
}

@article{Arkani-Hamed:1998jmv,
    author = "Arkani-Hamed, Nima and Dimopoulos, Savas and Dvali, G. R.",
    title = "{The Hierarchy problem and new dimensions at a millimeter}",
    eprint = "hep-ph/9803315",
    archivePrefix = "arXiv",
    reportNumber = "SLAC-PUB-7769, SU-ITP-98-13",
    doi = "10.1016/S0370-2693(98)00466-3",
    journal = "Phys. Lett. B",
    volume = "429",
    pages = "263--272",
    year = "1998"
}

@article{Antoniadis:1998ig,
    author = "Antoniadis, Ignatios and Arkani-Hamed, Nima and Dimopoulos, Savas and Dvali, G. R.",
    title = "{New dimensions at a millimeter to a Fermi and superstrings at a TeV}",
    eprint = "hep-ph/9804398",
    archivePrefix = "arXiv",
    reportNumber = "SLAC-PUB-7801, SU-ITP-98-28, CPTH-S608-0498, IC-98-39",
    doi = "10.1016/S0370-2693(98)00860-0",
    journal = "Phys. Lett. B",
    volume = "436",
    pages = "257--263",
    year = "1998"
}

@article{Arkani-Hamed:1998sfv,
    author = "Arkani-Hamed, Nima and Dimopoulos, Savas and Dvali, G. R.",
    title = "{Phenomenology, astrophysics and cosmology of theories with submillimeter dimensions and TeV scale quantum gravity}",
    eprint = "hep-ph/9807344",
    archivePrefix = "arXiv",
    reportNumber = "SLAC-PUB-7864, SU-ITP-98-142, IC-98-44",
    doi = "10.1103/PhysRevD.59.086004",
    journal = "Phys. Rev. D",
    volume = "59",
    pages = "086004",
    year = "1999"
}

@article{Arkani-Hamed:1998wuz,
    author = "Arkani-Hamed, Nima and Dimopoulos, Savas and Dvali, G. R. and March-Russell, John",
    title = "{Neutrino masses from large extra dimensions}",
    eprint = "hep-ph/9811448",
    archivePrefix = "arXiv",
    reportNumber = "SLAC-PUB-8014, SU-ITP-98-64",
    doi = "10.1103/PhysRevD.65.024032",
    journal = "Phys. Rev. D",
    volume = "65",
    pages = "024032",
    year = "2001"
}

@article{Dvali:2016uhn,
    author = "Dvali, Gia and Funcke, Lena",
    title = "{Small neutrino masses from gravitational \ensuremath{\theta}-term}",
    eprint = "1602.03191",
    archivePrefix = "arXiv",
    primaryClass = "hep-ph",
    reportNumber = "MPP-2016-277, LMU-ASC-31-16",
    doi = "10.1103/PhysRevD.93.113002",
    journal = "Phys. Rev. D",
    volume = "93",
    number = "11",
    pages = "113002",
    year = "2016"
}

@article{Dvali:2009ne,
    author = "Dvali, Gia and Redi, Michele",
    title = "{Phenomenology of $10^{32}$ Dark Sectors}",
    eprint = "0905.1709",
    archivePrefix = "arXiv",
    primaryClass = "hep-ph",
    reportNumber = "CERN-PH-TH-2009-054, MPP-2009-45",
    doi = "10.1103/PhysRevD.80.055001",
    journal = "Phys. Rev. D",
    volume = "80",
    pages = "055001",
    year = "2009"
}

@article{Dvali:2007hz,
    author = "Dvali, Gia",
    title = "{Black Holes and Large N Species Solution to the Hierarchy Problem}",
    eprint = "0706.2050",
    archivePrefix = "arXiv",
    primaryClass = "hep-th",
    doi = "10.1002/prop.201000009",
    journal = "Fortsch. Phys.",
    volume = "58",
    pages = "528--536",
    year = "2010"
}

@article{Dvali:2007wp,
    author = "Dvali, Gia and Redi, Michele",
    title = "{Black Hole Bound on the Number of Species and Quantum Gravity at LHC}",
    eprint = "0710.4344",
    archivePrefix = "arXiv",
    primaryClass = "hep-th",
    doi = "10.1103/PhysRevD.77.045027",
    journal = "Phys. Rev. D",
    volume = "77",
    pages = "045027",
    year = "2008"
}

@article{Ettengruber:2022pxf,
    author = "Ettengruber, Manuel",
    title = "{Neutrino physics in TeV scale gravity theories}",
    eprint = "2206.00034",
    archivePrefix = "arXiv",
    primaryClass = "hep-ph",
    doi = "10.1103/PhysRevD.106.055028",
    journal = "Phys. Rev. D",
    volume = "106",
    number = "5",
    pages = "055028",
    year = "2022"
}

@article{Dvali:1999cn,
    author = "Dvali, G. R. and Smirnov, Alexei Yu.",
    title = "{Probing large extra dimensions with neutrinos}",
    eprint = "hep-ph/9904211",
    archivePrefix = "arXiv",
    reportNumber = "NYU-TH-99-3-03",
    doi = "10.1016/S0550-3213(99)00574-X",
    journal = "Nucl. Phys. B",
    volume = "563",
    pages = "63--81",
    year = "1999"
}

@article{Machado:2011jt,
    author = "Machado, P. A. N. and Nunokawa, H. and Zukanovich Funchal, R.",
    title = "{Testing for Large Extra Dimensions with Neutrino Oscillations}",
    eprint = "1101.0003",
    archivePrefix = "arXiv",
    primaryClass = "hep-ph",
    doi = "10.1103/PhysRevD.84.013003",
    journal = "Phys. Rev. D",
    volume = "84",
    pages = "013003",
    year = "2011"
}

@article{Machado:2011kt,
    author = "Machado, P. A. N. and Nunokawa, H. and dos Santos, F. A. Pereira and Funchal, R. Zukanovich",
    title = "{Bulk Neutrinos as an Alternative Cause of the Gallium and Reactor Anti-neutrino Anomalies}",
    eprint = "1107.2400",
    archivePrefix = "arXiv",
    primaryClass = "hep-ph",
    doi = "10.1103/PhysRevD.85.073012",
    journal = "Phys. Rev. D",
    volume = "85",
    pages = "073012",
    year = "2012"
}

@article{Girardi:2014gna,
    author = "Girardi, I. and Meloni, D.",
    title = "{Constraining new physics scenarios in neutrino oscillations from Daya Bay data}",
    eprint = "1403.5507",
    archivePrefix = "arXiv",
    primaryClass = "hep-ph",
    reportNumber = "RM3-TH-14-3, SISSA-14-2014-FISI",
    doi = "10.1103/PhysRevD.90.073011",
    journal = "Phys. Rev. D",
    volume = "90",
    number = "7",
    pages = "073011",
    year = "2014"
}

@article{Rodejohann:2014eka,
    author = "Rodejohann, Werner and Zhang, He",
    title = "{Signatures of Extra Dimensional Sterile Neutrinos}",
    eprint = "1407.2739",
    archivePrefix = "arXiv",
    primaryClass = "hep-ph",
    doi = "10.1016/j.physletb.2014.08.035",
    journal = "Phys. Lett. B",
    volume = "737",
    pages = "81--89",
    year = "2014"
}

@article{Berryman:2016szd,
    author = "Berryman, Jeffrey M. and de Gouv\^ea, Andr\'e and Kelly, Kevin J. and Peres, O. L. G. and Tabrizi, Zahra",
    title = "{Large, Extra Dimensions at the Deep Underground Neutrino Experiment}",
    eprint = "1603.00018",
    archivePrefix = "arXiv",
    primaryClass = "hep-ph",
    doi = "10.1103/PhysRevD.94.033006",
    journal = "Phys. Rev. D",
    volume = "94",
    number = "3",
    pages = "033006",
    year = "2016"
}

@article{Carena:2017qhd,
    author = "Carena, Marcela and Li, Ying-Ying and Machado, Camila S. and Machado, Pedro A. N. and Wagner, Carlos E. M.",
    title = "{Neutrinos in Large Extra Dimensions and Short-Baseline $\nu_e$ Appearance}",
    eprint = "1708.09548",
    archivePrefix = "arXiv",
    primaryClass = "hep-ph",
    reportNumber = "FERMILAB-PUB-17-338-T",
    doi = "10.1103/PhysRevD.96.095014",
    journal = "Phys. Rev. D",
    volume = "96",
    number = "9",
    pages = "095014",
    year = "2017"
}

@article{Stenico:2018jpl,
    author = "Stenico, G. V. and Forero, D. V. and Peres, O. L. G.",
    title = "{A Short Travel for Neutrinos in Large Extra Dimensions}",
    eprint = "1808.05450",
    archivePrefix = "arXiv",
    primaryClass = "hep-ph",
    doi = "10.1007/JHEP11(2018)155",
    journal = "JHEP",
    volume = "11",
    pages = "155",
    year = "2018"
}

@article{Arguelles:2019xgp,
    author = {Arg\"uelles, C. A. and others},
    title = "{New opportunities at the next-generation neutrino experiments I: BSM neutrino physics and dark matter}",
    eprint = "1907.08311",
    archivePrefix = "arXiv",
    primaryClass = "hep-ph",
    reportNumber = "FERMILAB-FN-1079-T, Reports on Progress in Physics, Volume 83, Number 12",
    doi = "10.1088/1361-6633/ab9d12",
    journal = "Rept. Prog. Phys.",
    volume = "83",
    number = "12",
    pages = "124201",
    year = "2020"
}

@article{DUNE:2020fgq,
    author = "Abi, B. and others",
    collaboration = "DUNE",
    title = "{Prospects for beyond the Standard Model physics searches at the Deep Underground Neutrino Experiment}",
    eprint = "2008.12769",
    archivePrefix = "arXiv",
    primaryClass = "hep-ex",
    reportNumber = "FERMILAB-PUB-20-459-LBNF-ND",
    doi = "10.1140/epjc/s10052-021-09007-w",
    journal = "Eur. Phys. J. C",
    volume = "81",
    number = "4",
    pages = "322",
    year = "2021"
}

@article{Basto-Gonzalez:2021aus,
    author = "Basto-Gonzalez, V. S. and Forero, D. V. and Giunti, C. and Quiroga, A. A. and Ternes, C. A.",
    title = "{Short-baseline oscillation scenarios at JUNO and TAO}",
    eprint = "2112.00379",
    archivePrefix = "arXiv",
    primaryClass = "hep-ph",
    doi = "10.1103/PhysRevD.105.075023",
    journal = "Phys. Rev. D",
    volume = "105",
    number = "7",
    pages = "075023",
    year = "2022"
}

@inproceedings{Arguelles:2022xxa,
    author = {Arg\"uelles, C. A. and others},
    title = "{Snowmass White Paper: Beyond the Standard Model effects on Neutrino Flavor}",
    booktitle = "{2022 Snowmass Summer Study}",
    eprint = "2203.10811",
    archivePrefix = "arXiv",
    primaryClass = "hep-ph",
    month = "3",
    year = "2022"
}

@article{Arkani-Hamed:2016rle,
    author = "Arkani-Hamed, Nima and Cohen, Timothy and D'Agnolo, Raffaele Tito and Hook, Anson and Kim, Hyung Do and Pinner, David",
    title = "{Solving the Hierarchy Problem at Reheating with a Large Number of Degrees of Freedom}",
    eprint = "1607.06821",
    archivePrefix = "arXiv",
    primaryClass = "hep-ph",
    doi = "10.1103/PhysRevLett.117.251801",
    journal = "Phys. Rev. Lett.",
    volume = "117",
    number = "25",
    pages = "251801",
    year = "2016"
}

@article{Ettengruber:2024fcq,
    author = "Ettengruber, Manuel and Zander, Alan and Eller, Philipp",
    title = "{Testing the number of neutrino species with a global fit of neutrino data}",
    eprint = "2402.00490",
    archivePrefix = "arXiv",
    primaryClass = "hep-ph",
    doi = "10.1103/PhysRevD.109.095016",
    journal = "Phys. Rev. D",
    volume = "109",
    number = "9",
    pages = "095016",
    year = "2024"
}

@article{Bajc:2001fe,
    author = "Bajc, Borut and Senjanovic, Goran and Vissani, Francesco",
    editor = "Horv\'ath, Dezs\~o and L\'evai, P\'eter and Patk\'os, Andr\'as",
    title = "{How neutrino and charged fermion masses are connected within minimal supersymmetric SO(10)}",
    eprint = "hep-ph/0110310",
    archivePrefix = "arXiv",
    doi = "10.22323/1.007.0198",
    journal = "PoS",
    volume = "HEP2001",
    pages = "198",
    year = "2001"
}

@article{Brahmachari:1997cq,
    author = "Brahmachari, B. and Mohapatra, R. N.",
    title = "{Unified explanation of the solar and atmospheric neutrino puzzles in a minimal supersymmetric SO(10) model}",
    eprint = "hep-ph/9710371",
    archivePrefix = "arXiv",
    reportNumber = "UMD-PP-98-049",
    doi = "10.1103/PhysRevD.58.015001",
    journal = "Phys. Rev. D",
    volume = "58",
    pages = "015001",
    year = "1998"
}

@article{Aulakh:2000sn,
    author = "Aulakh, Charanjit S. and Bajc, Borut and Melfo, Alejandra and Rasin, Andrija and Senjanovic, Goran",
    title = "{SO(10) theory of R-parity and neutrino mass}",
    eprint = "hep-ph/0004031",
    archivePrefix = "arXiv",
    reportNumber = "NYU-TH-00-03-08",
    doi = "10.1016/S0550-3213(00)00721-5",
    journal = "Nucl. Phys. B",
    volume = "597",
    pages = "89--109",
    year = "2001"
}

@article{Dvali:1996wh,
    author = "Dvali, G. R. and Pokorski, Stefan",
    title = "{Naturally light Higgs doublet in the spinor representation of SUSY SO(10)}",
    eprint = "hep-ph/9601358",
    archivePrefix = "arXiv",
    reportNumber = "CERN-TH-96-11, MPI-PHT-96-4",
    doi = "10.1016/0370-2693(96)00357-7",
    journal = "Phys. Lett. B",
    volume = "379",
    pages = "126--134",
    year = "1996"
}

@article{Aulakh:1982sw,
    author = "Aulakh, C. S. and Mohapatra, Rabindra N.",
    title = "{Implications of Supersymmetric SO(10) Grand Unification}",
    reportNumber = "CCNY-HEP-82-4-REV, CCNY-HEP-82-4",
    doi = "10.1103/PhysRevD.28.217",
    journal = "Phys. Rev. D",
    volume = "28",
    pages = "217",
    year = "1983"
}

@article{Aulakh:2003kg,
    author = "Aulakh, Charanjit S. and Bajc, Borut and Melfo, Alejandra and Senjanovic, Goran and Vissani, Francesco",
    title = "{The Minimal supersymmetric grand unified theory}",
    eprint = "hep-ph/0306242",
    archivePrefix = "arXiv",
    doi = "10.1016/j.physletb.2004.03.031",
    journal = "Phys. Lett. B",
    volume = "588",
    pages = "196--202",
    year = "2004"
}

@article{Clark:1982ai,
    author = "Clark, T. E. and Kuo, Tzee-Ke and Nakagawa, N.",
    title = "{A SO(10) SUPERSYMMETRIC GRAND UNIFIED THEORY}",
    reportNumber = "PURD-TH-82-7",
    doi = "10.1016/0370-2693(82)90507-X",
    journal = "Phys. Lett. B",
    volume = "115",
    pages = "26--28",
    year = "1982"
}

@article{Weinberg:1979sa,
    author = "Weinberg, Steven",
    title = "{Baryon and Lepton Nonconserving Processes}",
    reportNumber = "HUTP-79-A050",
    doi = "10.1103/PhysRevLett.43.1566",
    journal = "Phys. Rev. Lett.",
    volume = "43",
    pages = "1566--1570",
    year = "1979"
}

@article{Minkowski:1977sc,
    author = "Minkowski, Peter",
    title = "{$\mu \to e\gamma$ at a Rate of One Out of $10^{9}$ Muon Decays?}",
    reportNumber = "Print-77-0182 (BERN)",
    doi = "10.1016/0370-2693(77)90435-X",
    journal = "Phys. Lett. B",
    volume = "67",
    pages = "421--428",
    year = "1977"
}

@article{Gell-Mann:1979vob,
    author = "Gell-Mann, Murray and Ramond, Pierre and Slansky, Richard",
    title = "{Complex Spinors and Unified Theories}",
    eprint = "1306.4669",
    archivePrefix = "arXiv",
    primaryClass = "hep-th",
    reportNumber = "PRINT-80-0576",
    journal = "Conf. Proc. C",
    volume = "790927",
    pages = "315--321",
    year = "1979"
}

@article{Yanagida:1980xy,
    author = "Yanagida, Tsutomu",
    title = "{Horizontal Symmetry and Masses of Neutrinos}",
    reportNumber = "TU-80-208",
    doi = "10.1143/PTP.64.1103",
    journal = "Prog. Theor. Phys.",
    volume = "64",
    pages = "1103",
    year = "1980"
}

@article{Mohapatra:1979ia,
    author = "Mohapatra, Rabindra N. and Senjanovic, Goran",
    title = "{Neutrino Mass and Spontaneous Parity Nonconservation}",
    reportNumber = "MDDP-TR-80-060, MDDP-PP-80-105, CCNY-HEP-79-10",
    doi = "10.1103/PhysRevLett.44.912",
    journal = "Phys. Rev. Lett.",
    volume = "44",
    pages = "912",
    year = "1980"
}

@inproceedings{Mohapatra:2004zh,
    author = "Mohapatra, R. N.",
    title = "{Seesaw mechanism and its implications}",
    booktitle = "{SEESAW25: International Conference on the Seesaw Mechanism and the Neutrino Mass}",
    eprint = "hep-ph/0412379",
    archivePrefix = "arXiv",
    doi = "10.1142/9789812702210_0003",
    pages = "29--44",
    month = "12",
    year = "2004"
}

\end{document}